\documentstyle[12pt]{elsart}
\begin{document}
\begin{frontmatter}
\title{Geometrical approach to the thermodynamical theory of phase transitions
of the second kind}
\author{A.K. Kanyuka}
\address{Institute of Metal Physics, Vernadsky,36, Kiev, 252680, Ukraine}
\author{V.S. Glukhov\thanksref{BDAMZ}}
\address{CSSA, ERL 306, Stanford University, Stanford, CA, 94305, U.S.A.}
\thanks[BDAMZ]{To whom all the correspondence should be addressed. Email:
glukhov@flare.stanford.edu}
\begin{abstract}
Geometrical approach to the phenomenological theory of phase transitions of
the second kind at constant pressure $P$ and variable temperature $T$  is
proposed. Equilibrium states of a
system at zero external field and fixed $P$ and $T$ are described by points
in three-dimensional space with coordinates $\eta$,
the order parameter, $T$, the temperature and $\phi$, the thermodynamic
potential.
These points form the so-called zero field  curve in the $(\eta, T, \phi)$
space. Its branch point
coincides with the critical point of the system. The small parameter
of the theory is the distance from the critical point along  the
zero-field curve. It is emphasized that no explicit functional dependency of
$\phi$ on $\eta$ and $T$ is imposed.

It is shown that using $(\eta, T, \phi)$ space one cannot overcome well-known
difficulties of
the Landau theory of phase transitions and describe non-analytical
behavior of real systems in the vicinity of the critical point. This
becomes possible only if one increases the dimensionality of the
space, taking into account the dependency of the thermodynamic
potential not only on  $\eta$ and $T$, but also on near (local) order
parameters $\lambda_{i}$. In this case under certain conditions it is
possible to describe  anomalous increase of the specific heat when the
temperature of the system approaches the critical point from above as
well as from below the critical temperature $T_{c}$.
\end{abstract}
\end{frontmatter}

\section{Introduction}

In describing phase transitions occurring in various systems at constant
pressure $P$
and variable temperature $T$ (or vice versa) one usually utilizes Gibbs'
thermodynamic potential $\Phi=E-TS+PV$, where $E$ is the internal
energy, $S$ is the entropy and $V$ is the volume \cite{Landau,Stanley}. Here we
shall consider phase transitions of the second kind at
constant pressure. In this case $\Phi$ as a function of the
generalized order parameter $\eta$ has one minimum at $\eta=0$ if $T >
T_c$, whereas if $T < T_c$ minimum of $\Phi$ is achieved at finite
values of $\eta=\eta_{*}(T)$.

In many cases $\Phi$ is an even function of the order parameter:
\begin{equation}\label{1}
\Phi(\eta,T,P)=\Phi(-\eta,T,P).
\end{equation}
For example, in ferromagnets $\Phi$ would not change if one
simultaneously reversed signs of the order parameter $\eta= M/M_{0}$ (
$M$ is the magnetic moment) and the external field $h=(\partial
\Phi/\partial M)_{P,T}$ \cite{Stanley,Vonsovskii}. The same is valid for
ferroelectrics. In the case of antiferromagnets \cite{Vonsovskii} and alloys
\cite{Krivoglaz}
reversing the sign of $\eta$ is reduced  to renaming equivalent sub-lattices.
In this article we shall consider the systems for which eq. (\ref{1})
is valid. The sketch of the dependency of $\Phi$ on $\eta$ and $T$ is shown on
Fig.1.

In the absence of the external field and $T <T_c$ homogeneous
equilibrium states corresponding to a given sign of $\eta$
 are achievable only in a system of an infinite size. In real systems of
a finite size at $T<T_c$ the homogeneous phase is unstable and the
equilibrium state breaks onto domains corresponding to $\eta_{*} >0$
and $\eta_{*} < 0$. The problem about equilibrium dimensions and
shapes of domains can be solved only in a theory taking into account
microscopic interactions in the system as well as the dimension and
the shape of the whole system. The thermodynamical approach does not allow to
describe heterogeneous states of a finite
system. Our goal is to establish general relations between proprieties
of $\Phi(\eta,T)$ and the behavior of basic thermodynamical
parameters of the system, i.e. the order parameter $\eta_{*}$, the
isobaric specific heat $C_{P}$ and the isothermal susceptibility
$\chi_{T}$, in the vicinity of the critical point.

The critical behavior of $\eta_{*}, C_{P}, \chi_{T}$ in the zero external
field can be expressed as
\begin{eqnarray}\label{2}
C_{P} &=& -T \frac{\partial^{2} \Phi}{\partial T^2} = a_{1}
(T-T_{c})^{-\alpha_1} + a_{2} (T-T_{c})^{-\alpha_2} \cdots,(\alpha_{1}
> \alpha_{2}) \nonumber \\
&& \\
\chi_{T}^{-1} &\sim& \left(\frac{\partial^{2} \Phi}{\partial
\eta^2}\right)_{T,P}
\propto (T-T_{c})^{\gamma}  \nonumber
\end{eqnarray}
at $T>T_c$ and as
\begin{eqnarray}\label{3}
C_{P} &=& a_{1}^{\prime}
(T-T_{c})^{-\alpha_{1}^{\prime}} + a_{2}^{\prime}
(T-T_{c})^{-\alpha_{2}^{\prime}} \cdots, (\alpha_{1}^{\prime}
> \alpha_{2}^{\prime}), \nonumber \\
\chi_{T}^{-1}&\propto& (T-T_{c})^{\gamma^{\prime}},\\
\eta_{*} &\propto& (T_{c}-T)^\beta  \nonumber
\end{eqnarray}
at $T<T_c$. Terms containing $\alpha_{2}$ and $\alpha_{2}^\prime$ are
necessary only when $\alpha_{1}=\alpha_{1}^{\prime}=0$. In that case
indexes $\alpha_{2}, \alpha_{2}^{\prime}$ describe critical behavior
of $dC_{P}/dT$ at $T \rightarrow T_{c} \pm 0$.

In order to characterize the system in the ``strong'' external field at
$T=T_{c}$ in the vicinity of the critical point one introduces
critical indexes $\delta$ and $\epsilon$:
\begin{eqnarray}\label{4}
\Phi(\eta, T_{c}) &\propto& \eta^{\delta+1},\nonumber \\
h &=& \frac{\partial \Phi}{\partial \eta} \propto \eta^{\delta},\\
C_{P} &\propto& h^{-\epsilon} \propto \eta^{-\delta \epsilon} \nonumber,
\end{eqnarray}
where $h$ is the generalized external field \cite{Landau}. For ferromagnets
and ferroelectrics indexes $\delta, \epsilon$ characterize real behavior of the
system
for $h \neq 0$, whereas for antiferromagnets and alloys they give
formal characteristics of $\Phi(\eta, T)$ in the vicinity of the
critical point.

Beside the indexes mentioned above one introduces also indexes
describing the behavior of the pair correlation function in the
vicinity of the critical point. However, the very concept of pair correlation
function is out of
the scope of our thermodynamical approach. Therefore, we
shall not discuss the corresponding critical indexes.

At first glance, the difference $T-T_{c}$ may play the role of a
convenient small
parameter when one considers phase transitions at constant pressure.
However, it is easy to prove that this is not quite true. The
definition of the isobaric specific heat is $C_{P}= -T (\partial^2
\Phi/ \partial T^2)_{P}$. On differentiating $\Phi$ twice by temperature we
obtain:
\begin{equation}
\left(\frac{\partial^2 \Phi}{ \partial T^2} \right)_{P} =\frac{ \partial^2
\Phi}{ \partial T^2} +
2 \frac{\partial^2 \Phi}{\partial \eta \partial T} \left(\frac{d
\eta}{d T}\right)_{P} + \frac{\partial^2 \Phi}{\partial \eta^2}
\left(\frac{d \eta}{d T}\right)_{P}^{2} + \frac{\partial
\Phi}{\partial \eta} \left(\frac{d^2 \eta}{d T^2}\right)_{P}, \label{5}
\end{equation}
\begin{equation}
\frac{\partial^2 \Phi}{\partial \eta \partial T} + \frac{\partial^2
\Phi}{\partial \eta^2} \left(\frac{d \eta}{d T}\right)_{P} =0. \label{6}
\end{equation}
Combining eqs. (\ref{5}), (\ref{6}) and taking into account that the
last term in eq. (\ref{6}) is identically zero in the case of the zero
external field, we obtain:
\begin{equation}
\left(\frac{\partial^2 \Phi}{ \partial T^2}\right)_{P} =\frac{ \partial^2
\Phi}{
\partial T^2} - \frac{ (\frac{\partial^2 \Phi}{\partial \eta \partial T})^2}{
\frac{\partial^2 \Phi}{\partial \eta^2}}. \label{7}
\end{equation}

As one can see from Fig.1  the derivative $
\partial^2 \Phi / \partial \eta^2$ is positive when $T>T_c$, $\eta=0$
and negative when when $T<T_c$,$\eta=0$. Therefore \footnote{Later we
will show this rigorously}, in the critical
point $\partial^2 \Phi / \partial \eta^2=0$. From the other hand, if
$T>T_c$, then from the condition $\eta=0$ and eq. (\ref{1}) it follows
that $\partial^2 \Phi/\partial \eta \partial T \equiv 0$. Therefore,
the second term in eq. (\ref{7}) is zero identically if $T \rightarrow
T_{c} +0$. For $T \rightarrow T_{c} -0$ the condition $h=0$
corresponds to the values $\eta = \eta_{*} \neq 0$. In this case the
derivative $\partial^2 \Phi/\partial \eta \partial T$ has, generally, a
finite value, thus the second term in eq. (\ref{7}) is indefinite.
Even more inconvenient is the calculation of $(\partial^3 \Phi
/\partial T^3)_{P}$ which characterizes the behavior of $dC_{P}/dT$.
One can overcome these difficulties if one chooses such small
parameter of the theory of which the thermodynamical values depend
smoothly. In the following sections we will show the advantages of the
geometrical choice of the small parameter and explain how, in our
opinion, a realistic phenomenological theory of phase transitions of
the second kind can be built.

\section{Geometrical interpretation of phase transitions of the second
kind}

Let us express the dependency of the thermodynamic potential $\Phi$ on
the order parameter $\eta$ and the dimensionless temperature $\tau=
T/T_{c}$ in the form:
\begin{equation}
\phi=\Phi(\eta,\tau T_{c}) \label{8}.
\end{equation}

Eq. (\ref{8}) defines a surface in the space $(\eta, \tau, \phi)$. The
condition $h=0$
\begin{equation}
\Phi_{\eta}(\eta, \tau T_{c})=0 \label{9},
\end{equation}
where subscript $\eta$ denotes the appropriate derivative, defines a
curve on the surface. Eqs. (\ref{8}),(\ref{9}) and the condition
$\Phi_{\eta \eta}>0$ define equilibrium states of the system at $h=0$ and
constant pressure $P$.

Consider an arbitrary point on the curve (\ref{8}),(\ref{9}) with
coordinates $\eta^{\prime}, \tau^{\prime}, \phi^{\prime}$. On
differentiating eqs. (\ref{8}),(\ref{9}) by the length $s$ of the
curve $(ds^2 = d\eta^2 + d \tau^2 + d \phi^2)$ we obtain:
\begin{eqnarray}
&&\Phi_{\eta} \dot{\eta} + \Phi_{\tau} \dot{\tau} = \dot{\phi}, \label{10}\\
&&\Phi_{\eta \eta} \dot{\eta} + \Phi_{\eta \tau} \dot{\tau} =0,\label{11}
\end{eqnarray}
where dot denotes the derivative with respect to $s$;$\dot{\eta}$,
$\dot{\tau}$ and $\dot{\phi}$ satisfy the normalizing condition
\begin{equation}
\dot{\eta}^2 + \dot{\tau}^2 + \dot{\phi}^2 =1. \label{12}
\end{equation}
Eqs. (\ref{10}),(\ref{11}),(\ref{12}) determine
$\dot{\eta},\dot{\tau},\dot{\phi}$ in an arbitrary point on the curve $h=0$.
Further differentiation of these
equations allows us to determine the derivatives of $\eta,\tau,\phi$ of
any order.They  are expressed in terms of derivatives of
$\Phi(\eta,\tau T_{c})$ in the the point $\eta^{\prime},
\tau^{\prime}, \phi^{\prime}$. In the vicinity of this point the
increments of the variables can be expressed as
\begin{eqnarray}
&&\eta-\eta^{\prime}=\sum{ \frac{1}{n!} \frac{d^n
\eta}{ds^n}},\nonumber \\
&&\tau-\tau^{\prime}=\sum{ \frac{1}{n!} \frac{d^n
\tau}{ds^n}},\label{13} \\
&&\phi-\phi^{\prime}=\sum{ \frac{1}{n!} \frac{d^n
\phi}{ds^n}}.\nonumber
\end{eqnarray}
Eliminating $s$ from the expressions for $\phi$ and $\tau$ one can obtain the
expression for $\phi-\phi^{\prime}$ as a function of $\tau-\tau^{\prime}$ with
any given precision.

Eq. (\ref{1}) yields
\begin{eqnarray}
&&\frac{\partial^{k+l} \Phi}{\partial^{k} \eta \partial^{l} \tau} = 0, k=2m+1;
l,m = 0,1,2 ...,\label{14}
\end{eqnarray}
whereas for even $k$ these derivatives are, generally, finite. In this
case one of the branches of the zero-field curve (\ref{8}),(\ref{9})
corresponds to $\eta=0$, for the condition (\ref{9}) is then valid
identically.

{}From eqs. (\ref{11}), (\ref{14})  for an arbitrary point of $\eta=0$ branch
of the
zero-field curve  it follows that $\dot{\eta} =0$. On differentiating
eq. (\ref{11}) by $s$ and taking into account eq. (\ref{14}) and
$\dot{\eta} =0$ we find that $\ddot{\eta}$ is also equal to
zero. Further differentiations  show that in an arbitrary point of the
$\eta=0$ branch of the zero-field curve where $\Phi_{\eta \eta}(0,
\tau T_{c}) \neq 0$ the derivatives of $\eta$ of
any order are equal to zero. This is  the mere consequence of the fact
that the branch under consideration lies on a plane $(\tau, \phi)$ in
the space $(\eta,\tau,\phi)$.

And only in the point where
\begin{eqnarray}
&&\Phi_{\eta \eta}(0,T_{c}) =0 \label{15}
\end{eqnarray}
there can
be a solution $\dot{\eta} \neq 0$ that corresponds to branches
$\eta=\eta_{*} \neq 0$ of the zero-field curve \footnote{If the
functional form of $\Phi(\eta,\tau T_{c})$ is known then
eq. (\ref{15}) can be
used to calculate $T_c$.}. Therefore, eq. (\ref{15}) determines the
branch point of the zero-field curve. We shall consider the vicinity
of this point. All derivatives of $\Phi(\eta, \tau T_{c})$ shall be
considered as taken in the point $\eta=0, \tau=1$.

In the case when eq. (\ref{14}),(\ref{15}) are valid eq.
(\ref{11}) becomes an identity. Therefore, in order to calculate
$\dot{\eta}, \dot{\tau}, \dot{\phi}$ we must differentiate
eqs. (\ref{10}),(\ref{11}),(\ref{12}) one more time:
\begin{eqnarray}
&&\Phi_{\eta \eta} \dot{\eta}^2 + 2 \Phi_{\eta \tau} \dot{\eta} \dot{\tau}
+ \Phi_{\tau \tau} \dot{\tau}^2 + \Phi_{\eta} \ddot{\eta} +
\Phi_{\tau} \ddot{\tau} = \ddot{\phi} \label{16}, \\
&&\Phi_{\eta \eta \eta} \dot{\eta}^2 + 2 \Phi_{\eta \eta \tau} \dot{\eta}
\dot{\tau}
+ \Phi_{\eta \tau \tau} \dot{\tau}^2 + \Phi_{\eta \eta} \ddot{\eta} +
 \Phi_{\eta \tau} \ddot{\tau} = 0 \label{17}, \\
&&\dot{\eta}\ddot{\eta} + \dot{\tau}\ddot{\tau} +
\dot{\phi}\ddot{\phi}=0.\label{18}
\end{eqnarray}

Eqs. (\ref{10}),(\ref{11}),(\ref{12}),(\ref{16}),(\ref{17}),(\ref{18})
yield two solutions:
\begin{eqnarray}
&&\dot{\eta}_{0}=0, \dot{\tau}_{0}= \pm \frac{1}{\sqrt{1+\Phi_{\tau}^2}},
\dot{\phi}_{0}=\Phi_{\tau} \dot{\tau}_{0},
\nonumber\\
&& \label{19}\\
&& \ddot{\tau}_{0} = -\frac{\Phi_{\tau} \Phi_{\tau
\tau}}{(1+\Phi_{\tau}^2)^{2}}, \ddot{\phi}_{0}= \Phi_{\tau} \ddot{\tau}_{0} +
\Phi_{\tau \tau} \dot{\tau}_{0}^2 \nonumber
\end{eqnarray}
and
\begin{eqnarray}
&&\dot{\eta}_{*}= \pm 1,
\dot{\tau}_{*}=\dot{\phi}_{*}=\ddot{\eta}_{*}=0, \ddot{\phi}_{*} =
\Phi_{\tau} \ddot{\tau}_{*}. \label{20}
\end{eqnarray}

Solutions (\ref{19}), (\ref{20})  correspond to the branch $\eta=0$
and branches $\eta_{*} \neq 0$ respectively . Here and thereafter
subscripts $0$ and $*$ denote
that values are taken on and derivatives are taken along the branch
$\eta=0$ and branches $\eta_{*} \neq 0$ respectively.

\subsection{Critical behavior for $\tau >1$}

In order to obtain critical indexes let us differentiate
eqs. (\ref{16}), (\ref{17}) and (\ref{18}) one more time. Taking into
account eqs. (\ref{14}), (\ref{15}) and (\ref{19}) we obtain:
\begin{eqnarray}
&& \phi_{0}^{(3)}=\Phi_{\tau} \tau_{0}^{(3)} + 3 \Phi_{\tau \tau}
\dot{\tau_{0}} \ddot{\tau_{0}} + \Phi_{\tau \tau \tau}
\dot{\tau}_{0}^3,\\
&&\dot{\tau_{0}}\tau_{0}^{(3)} + \dot{\phi_{0}}
\phi_{0}^{(3)} + \ddot{\phi_{0}}^2 =0,\label{21} \\
&&\Phi_{\eta \eta \tau} \dot{\tau}_{0} \ddot{\eta}_{0}=0,\label{22}
\end{eqnarray}
where superscripts $(n)$ denote $n$th derivatives with respect to $s$.

Eq. (\ref{22}) yields $\ddot{\eta_{0}}=0$, which is, as we already
discussed, a simple consequence that on the branch under consideration
$\eta \equiv 0$.

The expansion series (\ref{13}) for the variables $\phi$ and $\tau$ up
to the terms $\propto s^3$ are given by
\begin{eqnarray}
&&\tau-1=\dot{\tau_{0}} s + \frac{1}{2} \ddot{\tau_{0}} s^2 +
\frac{1}{6} \tau_{0}^{(3)}s^3,\\
&&(\tau -1)^2 = \dot{\tau_{0}}^2 s^2 +
\dot{\tau_{0}} \ddot{\tau_{0}} s^3,\\
&&(\tau-1)^3= \tau_{0}^{(3)}s^3,\label{23}\\
&&\phi-\phi_{0}=\dot{\phi_{0}} s + \frac{1}{2} \ddot{\phi_{0}} s^2 +
\frac{1}{6} \phi_{0}^{(3)}s^3, \label{24}
\end{eqnarray}
where $\phi_{0}=\Phi(0,T_{c})$.

Substituting into eq. (\ref{24}) the expressions (\ref{19}) and
(\ref{21}) for the derivatives of $\phi$ and taking into account
eq. (\ref{23}) we obtain
\begin{eqnarray}
&&\phi - \phi_{0}= \Phi_{\tau} (\tau -1) + \frac{1}{2} \Phi_{\tau \tau}
(\tau -1)^{2} + \frac{1}{6} \Phi_{\tau \tau \tau} (\tau-1)^{3} \label{25},
\end{eqnarray}
which is exactly the same as if one wrote the expansion series for
$\Phi(\eta \equiv 0, \tau T_{c})$ in the vicinity of $\tau=1$.

Eq. (\ref{25}) yields
\begin{eqnarray}
&&\left( \frac{d^2 \phi}{d \tau^2} \right)_{0} = \Phi_{\tau \tau} + \Phi_{\tau
\tau \tau} (\tau-1). \label{26}
\end{eqnarray}

$\Phi_{\eta \eta}(0, \tau T_{c})$ may also be written as an expansion
series in $s$ :
\begin{eqnarray}
&&\Phi_{\eta \eta}(0, \tau T_{c})= (\dot{\Phi}_{\eta \eta})_{0} s +
\frac{1}{2} (\ddot{\Phi}_{\eta \eta})_{0} s^2 + \cdots \label{27}
\end{eqnarray}
It is easily seen that
\begin{eqnarray}
&&(\dot{\Phi}_{\eta \eta})_{0} = \Phi_{\eta \eta \tau} \dot{\tau}_{0} .
\label{28}
\end{eqnarray}
Therefore, from eqs. (\ref{23}), (\ref{27}) and (\ref{28}) one obtains
\begin{eqnarray}
&&\Phi_{\eta \eta}(0, \tau T_{c}) = \Phi_{\eta \eta \tau} (\tau -1). \label{29}
\end{eqnarray}

If $\tau \geq 1$ and $\eta = 0$ the conditions $\Phi_{\eta \eta} \geq
0, C_{P} > 0$ must be satisfied. Taking into account eqs. (\ref{2}),
(\ref{26}) and (\ref{29}) we find that these conditions demand
\begin{eqnarray}
&&\Phi_{\tau \tau} <0, \Phi_{\eta \eta \tau} >0. \label{30}
\end{eqnarray}
{}From eqs. (\ref{2}), (\ref{26}) and (\ref{29}) also follow the values
of the critical indexes:
\begin{equation}
\alpha_{1}=0, \alpha_{2}=-1, \gamma=1. \label{31}
\end{equation}
i.e. $C_{P}$ and its derivative with respect to temperature are finite
in the limit $T \rightarrow T_{c} + 0$.

\subsection{Critical behavior for $\tau < 1$}

On differentiating eqs. (\ref{16}), (\ref{17}) and (\ref{18}) and
taking into account eqs. (\ref{14}), (\ref{15}) and the solution
(\ref{20}) we obtain
\begin{eqnarray}
&&\phi_{*}^{(3)}= \Phi_{\tau} \tau_{*}^{(3)}, \label{32}\\
&&\ddot{\tau_{*}}= -\Phi_{\eta \eta \eta \eta} \left( 3 \Phi_{\eta \eta
\tau}\right)^{-1}, \label{33}\\
&&\dot{\eta_{*}} \eta_{*}^{(3)} + \ddot{\tau}_{*}^{2} (1 + \Phi_{\tau}^2
) =0. \label{34}
\end{eqnarray}

The condition $ \tau < 1$ requires that $\ddot{\tau}_{*} <
0$. According to eqs. (\ref{30}) and (\ref{33}) this is possible if
\begin{eqnarray}
&&\Phi_{\eta \eta \eta \eta} > 0, \label{35}
\end{eqnarray}
i.e. $\Phi(\eta,\tau T_{c})$ as a function of the order parameter
$\eta$ has a minimum in the critical point.

According to eqs. (\ref{20}) and (\ref{34}) on the branch $\eta_{*}
>0, \dot{\eta}_{*}=1$ the derivative $\eta^{(3)}$ is negative, whereas
on the branch $\eta_{*} <0, \dot{\eta}_{*}=-1$ it is positive.

Second differentiation of eqs. (\ref{16}), (\ref{17}) and (\ref{18})
yields
\begin{eqnarray}
&&\phi_{*}^{(4)}=\Phi_{\tau} \tau_{*}^{(4)} + 3 \Phi_{\tau \tau}
\ddot{\tau}_{*}^{2} + 3 \Phi_{\eta \eta \tau} \ddot{\tau}_{*},
\label{36}\\
&&\tau_{*}^{(3)} = \phi_{*}^{(3)} = \eta_{*}^{(4)} =0, \label{37}
\end{eqnarray}
where eqs. (\ref{14}), (\ref{15}), (\ref{20}), (\ref{32}) and (\ref{33})
are taken into account.

Third differentiation of eqs. (\ref{16}), (\ref{17}) and (\ref{18})
completes the system of equations necessary to calculate the
derivatives $\tau_{*}^{(4)}, \phi_{*}^{(4)}$, and $\eta_{*}^{(5)}$.

Further differentiating of eqs. (\ref{16}), (\ref{17}) and (\ref{18})
reveals that on the branches $\eta_{*} \neq 0$ even derivatives of
$\tau, \phi$ and odd derivatives of $\eta$ are finite. Different
branches $\eta_{*} >0$ and $\eta_{*}<0$ correspond to different signs
of $\eta_{*}^{(2n+1)}$.

Fourth differentiation of eq. (\ref{16}) yields
\begin{eqnarray}
\phi_{*}^{(6)}&=& \nonumber \\
&=&\Phi_{\tau} \tau_{*}^{(6)} + 15 \left( \Phi_{\tau
\tau} \ddot{\tau}_{*} + \Phi_{\eta eta \tau}\right) \tau_{*}^{(4)}+ \\
& &+
15 \Phi_{\tau \tau \tau} \ddot{\tau}_{*}^{3} + 45 \Phi_{\eta
\tau}^{(2,2)}\ddot{\tau}_{*}^{2} + 15 \Phi{\eta \tau}^{(4,1)}
\ddot{\tau}_{*} + \Phi_{\eta \tau}^{(6,0)} \nonumber, \label{38}
\end{eqnarray}
where we introduced the notation
\begin{eqnarray}
&& \Phi_{\eta \tau}^{(n,m)} = \frac{ \partial^{n+m} \Phi}{\partial
\eta^{n} \partial \tau^{m}}. \nonumber
\end{eqnarray}

Expansion series (\ref{13}) on the branches $\eta_{*} \neq 0$ can be
written as
\begin{eqnarray}
\eta &=& \pm s, \label{39}\\
\tau-1&=& \frac{1}{2!} \ddot{\tau}_{*}s^2 + \frac{1}{4!} \tau_{*}^{(4)}s^4
+ \frac{1}{6!} \tau_{*}^{(6)}s^6, \label{40}\\
\phi-1&=& \frac{1}{2!} \ddot{\phi}_{*}s^2 + \frac{1}{4!} \phi_{*}^{(4)}s^4
+ \frac{1}{6!} \phi_{*}^{(6)}s^6, \label{41}
\end{eqnarray}

Eq. (\ref{40}) gives for up to $s^6$
\begin{eqnarray}
(\tau-1)^2&=& \frac{1}{4} \ddot{\tau}_{*}^{2} s^4 + \frac{1}{24}
\ddot{\tau}_{*} \tau_{*}^{(4)} s^6,\\
(\tau-1)^3&=& \frac{1}{8}
\ddot{\tau}_{*}^{3}s^6 \label{42}.
\end{eqnarray}

On eliminating parameter $s$ from the series (\ref{41}) using
expressions (\ref{42}) we finally obtain
\begin{eqnarray}
\phi -\phi_{0}&=& \nonumber\\
& &\Phi_{\tau} (\tau-1) + \frac{1}{2} \left( \Phi_{\tau
\tau} - 3 \Phi_{\eta \eta \tau}^{2} \Phi_{\eta \eta \eta \eta}^{-1}
\right) (\tau-1)^2 +\\
& &+\frac{1}{6} \tilde{\Phi}_{\tau \tau \tau} (\tau
-1)^3, \nonumber \label{43}
\end{eqnarray}
where
\begin{eqnarray}
&&\tilde{\Phi}_{\tau \tau \tau}= \Phi_{\tau \tau \tau}+ 3 \Phi_{\eta
\tau}^{(2,2)} \ddot{\tau}_{*}^{-1} + \Phi_{\eta\tau}^{(4,1)}
\ddot{\tau}_{*}^{-2} + \frac{1}{15} \Phi_{\eta\tau}^{(6,0)}
\ddot{\tau}_{*}^{-3}. \label{44}
\end{eqnarray}
Hence,
\begin{eqnarray}
&&\left( \frac{d^2 \phi}{d \tau^2} \right)_{*} = \Phi_{\tau \tau} - 3
\Phi_{\eta \eta \tau}^{2} \Phi_{\eta \eta \eta \eta}^{-1} +
\tilde{\Phi}_{\tau \tau \tau} (\tau -1). \label{45}
\end{eqnarray}

Comparing the expressions (\ref{26}) and (\ref{45}) and taking into account
inequalities (\ref{30}) and (\ref{30}) we obtain
\begin{eqnarray}
&&\left( \frac{d^2 \phi}{d \tau^2} \right)_{*} < \left( \frac{d^2
\phi}{d \tau^2} \right)_{0} <0, \label{46}
\end{eqnarray}
i.e., if the temperature decreases $C_{P}$ undergoes finite positive jump
in the critical point.

{}From eqs. (\ref{39}) and (\ref{40}) it follows that
\begin{eqnarray}
&&\left| \eta_{*} \right| \propto (1-\tau)^{\frac{1}{2}}.\label{47}
\end{eqnarray}

Expansion series (\ref{27}) remains valid for the branches $\eta_{*}
\neq 0$. Therefore, according to eqs. (\ref{14}), (\ref{20}) and (\ref{33})
\begin{eqnarray}
\left(\frac{d \Phi_{\eta \eta}}{d s} \right)_{*}&=& \Phi_{\eta \eta
\eta} \dot{\eta}_{*} + \Phi_{\eta \eta \tau} \dot{\tau}_{*} =0,\nonumber \\
&& \label{48}\\
\left(\frac{d^{2} \Phi_{\eta \eta}}{d s^{2}} \right)_{*}&=&
\Phi_{\eta \eta \eta \eta} \dot{\eta}_{*}^{2} + \Phi_{\eta \eta \tau}
\ddot{\tau}_{*} = - 2 \Phi_{\eta \eta \tau} \ddot{\tau}_{*} \nonumber.
\end{eqnarray}
Thus,
\begin{eqnarray}
&&\Phi_{\eta \eta}(\eta_{*},\tau T_{c})= - \Phi_{\eta \eta \tau}
\ddot{\tau}_{*} s^{2}.\label{49}
\end{eqnarray}
And, finally, from (\ref{30}) and (\ref{40}) we obtain
\begin{eqnarray}
&&\Phi_{\eta \eta}(\eta_{*},\tau T_{c})=2 \Phi_{\eta \eta \tau}
(1-\tau) >0.\label{50}
\end{eqnarray}

Comparing expressions (\ref{3}), (\ref{45}), (\ref{47}) and (\ref{50})
one readily obtains the values of the critical indexes:
\begin{eqnarray}
\alpha_{1}^{\prime}=0, \alpha_{2}^{\prime} = -1, \beta=\frac{1}{2},
\gamma^{\prime} = 1. \label{51}
\end{eqnarray}

\subsection{Critical behavior in the presence of the external field, $\tau=1$}

Consider the proprieties of the system on the line $\tau=1$.
If the conditions (\ref{14}), (\ref{15}) and (\ref{35}) are satisfied
the expansion series of $\phi$ as a function of $\eta$ is simply
\begin{eqnarray}
&&\phi -\phi_{0}= \frac{1}{4!} \Phi_{\eta \eta \eta \eta} \eta^{4} +
\frac{1}{6!} \Phi_{\eta \tau}^{(6,0)} \eta^{6}. \label{52}
\end{eqnarray}
Second term can be neglected near the first one. According to
(\ref{4}) we then obtain $\delta=3$.

In order to determine the critical index $\epsilon$ it is necessary to
calculate  second derivative of $\phi$ with respect to $\tau$ along
the line of a finite external field
\begin{equation}
h=\Phi_{\eta}(\eta, \tau T_{c}) = {\rm const} \neq 0. \label{53}
\end{equation}

In this case the second term in the right hand side of eq. (\ref{5})
is not identically zero and we need the expression for the second
derivative of $\eta$ with respect to $\tau$ along the line (\ref{53}).
On differentiating twice eq. (\ref{53}) we obtain
\begin{eqnarray}
&& \Phi_{\eta \tau} + \Phi_{\eta \eta} \left( \frac{d \eta}{d \tau}
\right)_{h} =0,\label{54}\\
&& \Phi_{\eta \tau \tau} + 2 \Phi_{\eta \eta \tau} \left( \frac{d
\eta}{d \tau} \right)_{h} + \Phi_{\eta \eta \eta} \left(
\frac{d \eta}{d \tau} \right)_{h}^{2} + \Phi_{\eta \eta}
\left(\frac{d^2 \eta}{d \tau^2}\right)_{h}=0,\nonumber
\end{eqnarray}
where  partial derivatives are taken along the line (\ref{53}) of the
surface (\ref{8}).

Substituting obtained from eqs. (\ref{54}) expressions for the
derivatives of $\eta$ into eq. (\ref{5}) we obtain
\begin{eqnarray}
\left( \frac{d^2 \phi}{d \tau^2} \right)_{h} &=& \nonumber\\
&=& \Phi_{\tau \tau} -
\Phi_{\eta \tau}^{2} \Phi_{\eta \eta}^{-1} - \\
& &-\Phi_{\eta}
\left(\Phi_{\eta \tau \tau} \Phi_{\eta \eta}^{-1} -2 \Phi_{\eta \tau}
\Phi_{\eta \eta \tau} \Phi_{\eta \eta}^{-2} + \Phi_{\eta \tau}^{2}
\Phi_{\eta \eta \eta} \Phi_{\eta \eta}^{-3}\right) \nonumber. \label{55}
\end{eqnarray}

Expression (\ref{55}) is obtained for an arbitrary point of the
surface (\ref{8}). It is valid indeed for the line $\tau=1$ of this
surface. In the vicinity of the critical point the derivatives of
$\Phi_(\eta, \tau T_{c})$ can be expressed as the expansion series in
$\eta$. Taking into account (\ref{14}), (\ref{15}) and (\ref{35}) we
obtain
\begin{eqnarray}
&& \Phi_{\tau \tau}(\eta, T_{c})= \Phi_{\tau \tau} \eta^0,
\Phi_{\eta \eta \tau}(\eta, T_{c})=\Phi_{\eta \eta \tau} \eta^0,
\Phi_{\eta \tau}(\eta, T_{c})= \Phi_{\eta \eta \tau} \eta
,\nonumber \\
&& \Phi_{\eta \tau \tau}(\eta, T_{c})= \Phi_{\eta \tau}^{(2,2)}
\eta, \Phi_{\eta \eta \eta}(\eta, T_{c})= \Phi_{\eta \eta \eta
\eta} \eta ,\label{56} \\
&& \Phi_{\eta \eta}(\eta, T_{c})= \frac{1}{2} \Phi_{\eta \eta \eta
\eta} \eta^2 , \Phi_{\eta}(\eta, T_{c})= \frac{1}{6} \Phi_{\eta \eta
\eta \eta} \eta^3, \nonumber
\end{eqnarray}
where the derivatives in the right hand side are taken at $\eta=0,
\tau=1$.

Substituting expressions (\ref{56}) into eq. (\ref{55}) we can verify
that the right hand side of eq. (\ref{55}) in the limit $\eta
\rightarrow 0$ is finite. According to (\ref{4}) this case corresponds
to $\delta \epsilon =0$. Therefore,
\begin{equation}
\delta=3, \epsilon =0. \label{57}
\end{equation}

Obtained critical indexes (\ref{31}), (\ref{51}) and (\ref{57})
coincide with ones obtained in the frameworks of the Landau theory
\cite{Landau} except  the calculation  of $\alpha_{2}$ and
$\alpha_{2}^{\prime}$. This can be explained by the fact that here we
considered by a different approach essentially the same
situation as in \cite{Landau}, namely, the problem under constraints
(\ref{14}), (\ref{15}) and (\ref{35}).

\subsection{Tricritical point}

Let us consider the situation when beside eq. (\ref{15}) the following
conditions are satisfied:
\begin{equation}
\Phi_{\eta \eta \eta \eta}=0, \Phi_{\eta \tau}^{(6,0)} >0. \label{58}
\end{equation}

The derivatives (\ref{58}) do not give any contribution into eqs. (\ref{21}),
(\ref{22}),  (\ref{23}),  (\ref{24}),  (\ref{25}),  (\ref{26}),
(\ref{27}),  (\ref{28}), (\ref{29}) and (\ref{30}). Therefore, the
critical behavior on the branch $\eta=0, \tau \geq 1$ remains
unchanged and the values of critical indexes are given by (\ref{31}).
The critical behavior of the branches $\eta_{*} \neq 0$, however, is
completely different.

According to eqs. (\ref{20}), (\ref{33}),  (\ref{34}),  (\ref{36}) and
(\ref{58}) in this case
\begin{eqnarray}
&&\ddot{\tau}_{*}=\ddot{\phi}_{*}=\tau_{*}^{(3)}=0, \label{59} \\
&& \phi_{*}^{(4)} = \Phi_{\tau} \tau_{*}^{(4)}. \label{60}
\end{eqnarray}

In order to calculate fourth derivative of $\tau$ along the branch
$\eta_{*} \neq 0$ it is sufficient to differentiate eq. (\ref{17})
three times and take into account eqs. (\ref{14}), (\ref{15}),
(\ref{20}),  (\ref{30}),  (\ref{58}) and  (\ref{59}). The result is
\begin{equation}
\tau_{*}^{(4)} = -\frac{1}{5} \Phi_{\eta \tau}^{(6,0)} \Phi_{\eta \eta
\tau}^{-1} < 0.\label{61}
\end{equation}

Eq. (\ref{38}) has the form
\begin{equation}
\phi_{*}^{(6)} = \Phi_{\tau} \tau_{*}^{(6)} - 2 \Phi_{\eta
\tau}^{(6,0)}.\label{62}
\end{equation}

According to (\ref{59}) in the expansion series (\ref{40}) and
(\ref{41}) the quadratic in $s$ terms are missing. Therefore,
according to (\ref{40}), (\ref{41}),  (\ref{58}),  (\ref{59}),
(\ref{60}),  (\ref{61}) and   (\ref{62}) it follows that
\begin{equation}
\phi - \phi_0= \Phi_{\tau} (\tau-1) - \frac{4}{3} D (1-\tau)^{\frac{3}{2}},
\end{equation}
where
\begin{equation}
D = \left( \frac{15}{2} \Phi_{\eta \eta \tau}^{3}
 \left(\Phi_{\eta \tau}^{(6,0)}\right)^{-1} \right)^{\frac{1}{2}} >0.\label{63}
\end{equation}
Thus
\begin{equation}
\left( \frac{d^2 \phi}{d \tau^2} \right)_{*} = -D
(1-\tau)^{-\frac{1}{2}} \label{64}
\end{equation}

{}From eqs.  (\ref{39}),  (\ref{40}) and  (\ref{61}) we obtain
\begin{equation}
\left| \eta_{*} \right| \propto (1-\tau)^{\frac{1}{4}}. \label{65}
\end{equation}

According to (\ref{48}), (\ref{58}) and  (\ref{59}) first and second
derivatives  of $\Phi_{\eta \eta}$ along the branch $\eta_{*} \neq 0$
are equal to zero in the tricritical point. On differentiating
$\Phi_{\eta \eta}$ third and fourth time and taking into account
 (\ref{14}), (\ref{15}), (\ref{20}), (\ref{58}), (\ref{59}) and
(\ref{61}) we obtain:
\begin{eqnarray}
&&\left( \frac{d^3 \Phi_{\eta \eta }}{d s^3} \right)_{*} =0,\nonumber
\\
&& \label{66}\\
&&\left( \frac{d^4 \Phi_{\eta \eta }}{d s^4} \right)_{*} = \Phi_{\eta
\tau}^{(6,0)} \dot{\eta}_{*}^{4} + \Phi_{\eta \eta \tau}
\tau_{*}^{(4)}= - 4 \Phi_{\eta \eta \tau} \tau_{*}^{(4)}.\nonumber
\end{eqnarray}
Therefore, the expansion series for $\Phi_{\eta \eta}$ along the
branch $\eta_{*} \neq 0$ is given by
\begin{equation}\label{67}
\Phi_{\eta \eta}( \eta_{*}, \tau T_{c}) = -\frac{1}{3!} \Phi_{\eta \eta
\tau} \tau_{*}^{(4)} s^4.
\end{equation}
Hence,
\begin{equation}\label{68}
\Phi_{\eta \eta}( \eta_{*}, \tau T_{c}) = 4 \Phi_{\eta \eta \tau}
(1-\tau) >0.
\end{equation}

According to  (\ref{14}), (\ref{15}), (\ref{52}) and  (\ref{58}) on
line $\tau =1$ of the surface (\ref{8})
\begin{eqnarray}
\Phi_{\tau \tau}(\eta, T_{c}) &\propto& \Phi_{\eta \eta \tau}(\eta, T_{c})
\propto \eta^{0}, \nonumber\\
\Phi_{\eta \tau}(\eta, T_{c}) &\propto& \Phi_{\eta \tau \tau} (\eta, T_{c})
\propto \eta^{1}, \nonumber\\
\Phi_{\eta \eta \eta}(\eta, T_{c}) &\propto& \eta^{3} ,\label{69}\\
\Phi_{\eta \eta}(\eta, T_{c}) &\propto& \eta^{4},\nonumber\\
\Phi_{\eta}(\eta, T_{c}) &\propto& \eta^{5}. \nonumber
\end{eqnarray}

In this case eq. (55) yields
\begin{equation}
\left( \frac{d^2 \phi}{d \tau^2} \right)_{h} \propto \eta ^{-2} \label{70}.
\end{equation}

Therefore, according to (\ref{3}), (\ref{4}), (\ref{52}), (\ref{58}),
(\ref{64}), (\ref{65}), (\ref{68}) and (\ref{70}) the critical indexes for
$\tau \leq 1$ are
\begin{equation}
\alpha_{1}^{\prime}=\frac{1}{2}, \beta =\frac{1}{4}, \gamma^{\prime}
=1, \delta=5, \epsilon = \frac{2}{5}, \label{71}
\end{equation}
whereas for $\tau > 1$ the indexes are the same as in (\ref{31}).

Similar consideration can be applied to the case when the first
non-vanishing derivative  of $\Phi$ with respect to $\eta$ is of order
$2k$ ($k$--critical point). In this case the critical indexes for $\tau \leq 1$
are
\begin{equation}
\alpha_{1}^{\prime}=\frac{k-2}{k-1}, \beta =\frac{1}{2k-2}, \gamma^{\prime}
=1, \delta\
=2k-1, \epsilon = \frac{2k-4}{2k-1}. \label{72}
\end{equation}

Fig. 2 represents  the sketch of $C_{P}(T)$ for $k=2$ and $k > 2$. It
is known that neither Fig.2 nor the critical indexes (\ref{31}),
(\ref{72}) are in agreement with experimental data. For a majority of
real systems $1/3 \leq \beta \leq 1/2, 1 \leq \gamma \leq 4/3, 3 \leq
\delta \leq 5$ and $\alpha_{1}, \alpha_{2}, \epsilon$ do not deviate
significantly from zero. Specific heat $C_{P}$ anomalously increases as
one approaches $T_{c}$ from above as well as from below.

This character of phase transitions in real systems is usually
attributed to non-analyticity of $\Phi$ in the critical point,
particularly, to divergence of partial derivatives of $\Phi$. It it is so
then the realistic phenomenological theory in either Landau's or in
proposed here geometrical formulation will be impossible.

There is, however, one reason in favor of an ``analytical'' approach.
According to (\ref{15}), (\ref{20}), (\ref{58}) and (\ref{64}) on the
branch $\eta_{*} \neq 0$, which approaches the critical point
along the normal to the surface $(\phi, \tau)$, the derivative $(d^2
\phi /d \tau^2)_{*}$ may increase infinitely even if the partial derivatives
of $\Phi$ with respect to $\eta$ and $\tau$ remain finite. In order
that such anomalous increase of $d^2 \phi /d \tau^2$ be possible on
the branch $\eta =0, \tau \rightarrow 1+0$ it is necessary to ``take
this branch out'' of the surface $(\phi, \tau)$ and make it a spatial
curve. The feasible approach to this problem is to take into account
the dependency of the thermodynamic potential on additional internal
parameters  characterizing the configuration of the system undergoing
phase transition. Contrary to the order parameter $\eta$ these additional
parameters  must essentially depend on the temperature  above as well
as
below $T_{c}$.

\section{Dependency of the thermodynamic potential on generalized
correlation parameters}

A system undergoing phase transition is characterized by far as well
as near (local) order \cite{Vonsovskii,Krivoglaz}. The far order vanishes above
the
critical point, whereas correlations describing the local order are
finite and essentially depend on temperature both above and below the
critical point.

Attempts to create a consistent statistical theory allowing to take
into account correlations meet with obstacles of computational as well
as theoretical character. Contrary to this, a phenomenological
approach avoids the explicit choice of correlation parameters and the
calculation of the thermodynamic potential.
For the phenomenological
approach proposed in this section it is important only that the number
of distinguishable configurations of the system is limited and they can be
characterized by a limited number of correlation parameters
$\lambda_{i}, i=0,....N-1$.

Let the function $\Phi(\eta, \lambda_{i}, T,P)$ defines the dependency
of the thermodynamic potential on the degree of the far order $\eta$,
correlation parameters $\lambda_{i}$, temperature $T$ and pressure
$P$. Then
\begin{equation}
\phi=\Phi(\eta, \lambda_{i}, T, P) \label{73}
\end{equation}
defines a hypersurface in the $N+4$--dimensional space $(\eta,
\lambda_{i}, T,P,\phi)$. Equilibrium states of the system for given
$\eta, T, P$ are the points of the hypersurface where
\begin{equation}
\frac{\partial \Phi}{\partial \lambda_{i}} =0. \label{74}
\end{equation}
and the matrix
\begin{equation}
A_{0}=\left\|\frac{\partial^2 \Phi}{\partial \lambda_{i} \partial
\lambda_{j}} \right\| \label{75}
\end{equation}
is positively determined.

The zero-field line on the surface (\ref{73}) is defined by the conditions
\begin{equation}
\frac{\partial \Phi}{\partial \eta} =0, \frac{\partial \Phi}{\partial
\lambda_{i}}=0
\end{equation}
and matrix
\begin{equation}\label{77}
A_{1}= \left\| \begin{array}{cccc}
\frac{\partial^2 \Phi}{\partial \eta^2}&\frac{\partial^2 \Phi}{\partial \eta
\partial \lambda_{0}}&\frac{\partial^2 \Phi}{ \partial \eta \partial
\lambda_{1}}& \cdots\\
\frac{\partial^2 \Phi}{\partial \lambda_{0} \partial \eta}&\frac{\partial^2
\Phi}{\partial \lambda_{0}^2}&\frac{\partial^2 \Phi}{\partial \lambda_{0}
\lambda_{1}}& \cdots\\
\frac{\partial^2 \Phi}{\partial \lambda_{1} \partial \eta}&\frac{\partial^2
\Phi}{\partial \lambda_{1} \partial \lambda_{0}}&\frac{\partial^2
\Phi}{\partial \lambda_{1}^2}& \cdots\\
\vdots&\vdots&\vdots&\ddots
\end{array}
       \right\|
\end{equation}
is positively determined.

Let us differentiate $\Phi(\eta, \lambda_{i}, T,P)$ on T at $P=$const,
$\eta=$const:
\begin{equation}\label{78}
\left( \frac{\partial \Phi}{\partial T} \right)_{\eta, P}= \frac{\partial
\Phi}{\partial T} + \frac{\partial \Phi}{\partial \lambda_{i}} \left( \frac{d
\lambda_{i}}{d T} \right)_{P}.
\end{equation}
In eq. (\ref{78}) and thereafter the summation over the repeated index is
assumed.

The derivative in the left hand side of eq. (\ref{78})  is the entropy of the
system (up to the sign of the expression). Owing to (\ref{74}) for equilibrium
states of the system $S= - \partial \Phi / \partial T$. Let
$\lambda_{i}$ be a full set of correlation parameters in the sense that
$\eta$ and $\lambda_{i}$ give the complete description of the
configuration of the system. In this case $S$ and $\partial \Phi /
\partial T$ are functions of $\eta$ and $\lambda_{i}$ only and do not
depend explicitly on $T$.

On differentiating eqs. (\ref{74}) and (\ref{78}) with respect to $T$ we find:
\begin{equation}\label{79}
\left( \frac{\partial^2 \Phi}{\partial T^2} \right)_{\eta, P}=
\frac{\partial^2 \Phi}{\partial T^2} + 2 \frac{\partial^2
\Phi}{\partial T \partial \lambda_{i}} \left( \frac{d \lambda_{i}}{d
T} \right)_{P} + \frac{\partial^2 \Phi}{\partial \lambda_{i} \partial
\lambda_{j}} \left( \frac{d \lambda_{i}}{d T} \right)_{P}\left(
\frac{d \lambda_{j}}{d T} \right)_{P},
\end{equation}
\begin{equation}
\frac{\partial^2 \Phi}{\partial T \partial \lambda_{i}} +
\frac{\partial^2 \Phi}{\partial \lambda_{i} \partial \lambda_{j}}
\left(\frac{d \lambda_{j}}{d T} \right)_{P} =0.\label{80}
\end{equation}
Eqs. (\ref{79}) and (\ref{80}) yield
\begin{equation}\label{81}
\left( \frac{\partial^2 \Phi}{\partial T^2} \right)_{\eta,
P}=\frac{\partial^2 \Phi}{\partial T^2} - \frac{\partial^2
\Phi}{\partial \lambda_{i} \partial \lambda_{j}}\left( \frac{d \lambda_{i}}{d
T} \right)_{P}\left(\frac{d \lambda_{j}}{d T} \right)_{P}.
\end{equation}

Since $S$ does not depend explicitly on $T$ the first term in eq.
(\ref{81}) is zero. The second term is negative because the matrix
(\ref{75}) is positively determined. By definition, the derivative in
the left hand side of eq. (\ref{81}) in
the case of ferromagnets is (for up to the factor $-T$) the specific heat
$C_{M}$, i.e. the specific
heat at constant magnetic moment $M=M_{0}$ and pressure $P$.
Therefore, $C_{M} > 0$.

Eliminating from eqs. (\ref{79}) and (\ref{80}) derivatives
$(d \lambda_{i}/dT)_{\eta,P}$ we find:
\begin{equation}\label{82}
\left( \frac{\partial^2 \Phi}{\partial T^2} \right)_{\eta,P}=
\frac{\det A_{2}}{\det A_{0}},
\end{equation}
where the expression for $A_{2}$ can be  obtained from (\ref{77}) by
replacing $\eta \rightarrow T$.

Similarly, the following expression is also obtained:
\begin{equation}\label{83}
\left( \frac{\partial^2 \Phi}{\partial \eta^2}
\right)_{T,P}=\frac{\det A_{1}}{\det A_{0}}
\end{equation}

In order to determine the derivative $(\partial^2 \Phi/\partial
T^2)_{P}$ along the line of zero field on the surface (\ref{73}) it is
necessary to differentiate $\Phi(\eta, \lambda_{i}, T,P)$ twice by $T$
taking into account the dependency of $\eta$ and $\lambda_{i}$
on $T$.  Eliminating from obtained expressions $(d
\lambda_{i}/dT)_{P}$ we come up to the equation
\begin{equation}\label{84}
\left( \frac{\partial^2 \Phi}{\partial T^2} \right)_{P} = \left(
\frac{\partial^2 \Phi}{\partial T^2} \right)_{\eta,P} - \left(
\frac{\partial^2 \Phi}{\partial \eta^2} \right)_{T,P} \left( \frac{d
\eta}{d T} \right)_{P}^{2}.
\end{equation}

Derivatives (\ref{82}) and (\ref{83}) remain finite in the critical
point of the surface (\ref{73}) if the matrix $A_{0}$ remain
positively determined. According to Krivoglaz \cite{Krivoglaz2} in this case
the
values of critical indexes are the same as in the Landau theory,
namely, (\ref{31}) and (\ref{72}).

Before we shall discuss the situation when $\det A_{0}=0$ in the
critical point, let us note that $A_{0}$ is real and symmetric.
Therefore, it is always possible to find a diagonal representation of
this matrix by an appropriate choice of $\lambda_{i}$. Let then in the
critical point
\begin{equation}\label{85}
\frac{\partial^2 \Phi}{\partial \lambda_{i} \partial \lambda_{j}}=
\delta_{ij} \frac{\partial^2 \Phi}{\partial \lambda_{i}^2}.
\end{equation}

Then, taking into account $\partial^2 \Phi/\partial T^2 =0$, from eq.
(\ref{81}) we obtain
\begin{equation}\label{86}
\left( \frac{\partial^2 \Phi}{\partial T^2} \right)_{\eta,P}= -
\frac{\partial^2 \Phi}{\partial \lambda_{i}^2} \left(\frac{d
\lambda_{i}}{d T} \right)_{P}^2.
\end{equation}

Let us assume that in the critical point
\begin{equation}\label{87}
\frac{\partial^2 \Phi}{\partial \lambda_{0}^2}=0, \frac{\partial^2
\Phi}{\partial \lambda_{i}^2} > 0, i=1,2,\cdots N-1.
\end{equation}
Then, from eq. (\ref{80}) it follows that if one approaches the critical point
the derivative $(d \lambda_{0}/d T)_{p}$ may become infinite while the
derivatives of $\lambda_{i}$ $i=1,2,\dots,N-1$ remain finite. According to
 (\ref{86}) and (\ref{87}) eq. (\ref{84}) in the critical point becomes
\begin{eqnarray}
\left( \frac{\partial^2\Phi}{\partial T^2} \right)_{P}&=& \nonumber\\
&&- \lim_{T
\rightarrow T_{c}} \left( \left(\frac{\partial^2 \Phi}{\partial
\eta^2} \right)_{T,P} \left(\frac{d \eta }{d T} \right)_{P}^2 +
\frac{\partial^2 \Phi}{\partial \lambda_{0}^2} \left(\frac{d \lambda_{0}}{d T}
\right)_{P}^{2}  \right) -  \\
&&-\frac{\partial^2 \Phi}{\partial \lambda_{i}^2} \left(\frac{d \lambda_{i}}{d
T} \right)_{P}^{2}, (i>0) \nonumber \label{88}
\end{eqnarray}

Thus, it is ``critical variables'' $\eta$ and $\lambda_{0}$ which
may provide the infiniteness of the derivative (\ref{88}) and the
specific heat $C_{P}$ in the critical point, whereas the input of
other parameters is regular.

On the zero-field line for $T>T_{c}$ the equilibrium  values of $\eta$
and $(d\eta/dT)_{P}$ are zero identically. Therefore, for $T>T_{c}$
the first term in (\ref{88}) is zero. It is the second term that may lead to
anomalous increase of the specific heat at $T \rightarrow T_{c} +0$.

The conditions (\ref{85}) and (\ref{87}) greatly reduce the
dimensionality of the space where critical behavior is studied.
Substituting equilibrium values of ``regular'' $\lambda_{i}(\eta,
\lambda_{0}, \tau T_{c}), i>0 $ into eq. (\ref{73}) one obtains
\begin{equation}\label{89}
\phi=\Phi^{\prime}(\eta, \lambda_{0}, \tau T_{c}).
\end{equation}

Those points of surface (\ref{89}) in the four-dimensional space
$(\eta, \lambda_{0}, \tau, \phi)$ for which
\begin{equation}\label{90}
\frac{\partial \Phi^{\prime}}{\partial \eta}=0, \frac{\partial
\Phi^{\prime}}{\partial \lambda_{0}}=0, \frac{\partial^2
\Phi}{\partial \eta^2} >0,
\end{equation}
\begin{equation}\label{91}
\frac{\partial^2 \Phi^{\prime}}{\partial \eta^2}\frac{\partial^2
\Phi^{\prime}}{\partial \lambda_{0}^2}- \left( \frac{\partial^2
\Phi^{\prime}}{\partial \eta \partial \lambda_{0}} \right)^2 >0
\end{equation}
constitute the zero-field line.

Since we reduced the set of correlation parameters to one the entropy
$S=-\partial \Phi^{\prime}/\partial T$ of the system may explicitly
depend on $T$. Because of this, the derivatives of highest than the
first order of $\Phi^{\prime}$ by
$\tau$ may not be zero.

The critical behavior on the zero-field line and at $\tau=1$ in the
vicinity of the critical point will be determined by the order of
first non-vanishing derivatives
\begin{equation}\label{92}
\frac{\partial^{(2k)} \Phi}{\partial \eta^{(2k)}} >0,
\frac{\partial^{(2n)} \Phi}{\partial \lambda_{0}^{(2n)}} >0
\end{equation}

In forthcoming publications we will show the dependency of the
critical indexes on the parameters $k$ and $n$.

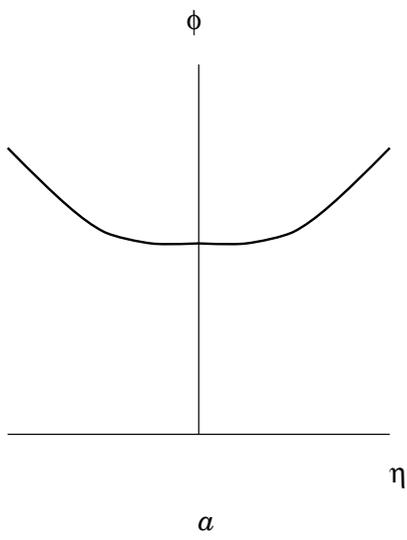
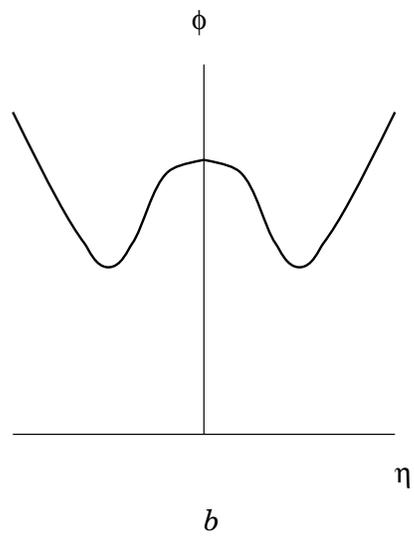
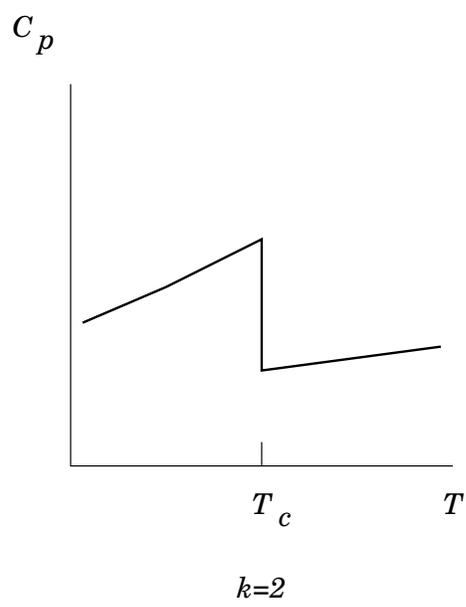
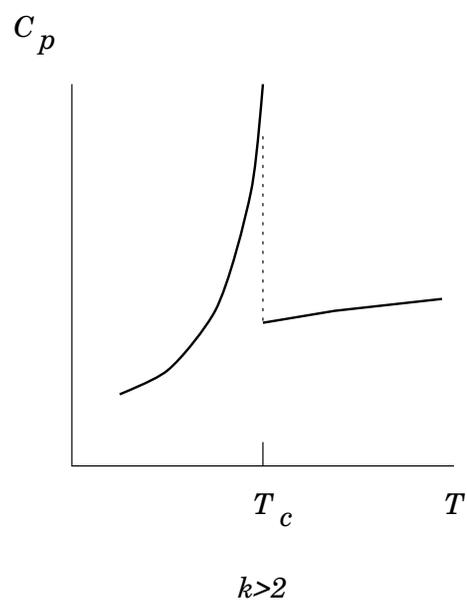
\end{document}